# Queering AI: Undoing the Self in the Algorithmic Borderlands

**(In review)**

Grace Leonora Turtle*

Delft University of Technology, Netherlands, g.l.turtle@tudelft.nl

Roy Bendor

Delft University of Technology, Netherlands, r.bendor@tudelft.nl

Elisa Giaccardi

Politecnico di Milano, Italy, elisa.giaccardi@polimi.it

Blazej Kotowski

Universitat Pompeu Fabra, Spain, blazej.kotowski@ufp.edu

This paper challenges fixed orientations towards the self in human-AI entanglements. It offers queering as a strategy to subvert the individuation and fixing of identities within algorithmic systems and the loss of futurity that it brings about. By exploring queerness, the paper examines how one's sense of self and futurity are interpellated within the algorithmic borderlands of human-AI entanglements. The study discusses an embodied experiment called "Undoing Gracia," a Digital Twin simulation where the first author Grace and their AI twins (Lex and Tortugi) interact within the fictional world of Gracia. The experiment probes into Grace's multifaceted subjectivities by conceiving themselves as interdependent entities evolving through their interactions within Gracia. The paper outlines the process of creating and implementing the simulation and examines how the agents co-perform and become-with alongside Gracia's making. The findings illuminate queer gestures for navigating human-AI entanglements in HCI research and practice, highlighting the importance of fluid identities in shaping human-AI relations.

---

[1] A Digital Twin, typically a simulation model, consists of a physical entity, its virtual counterpart, and the data and information flows that connect the two, enabling the virtual to reflect the physical [Grieves, 2005].

## 1 INTRODUCTION

Recent developments in Artificial Intelligence are blurring the lines between humanity and technology. As society becomes more entangled with AI at all levels—from individual humans to entire cities—questions arise about the future of human-computer interaction in research and practice [Frauenberger, 2019; Giaccardi et al. 2024]. Where does the human body end and the virtual one begin? How much of reality is mirrored, distorted, or even surpassed by a Digital Twin (DT)?[2] These (and similar) questions signal the dilemmas that come with integrating AI technologies into everyday life and underscore the need for researchers to explore new avenues for studying the evolving relationships and novel interactions between humans and computers.

Data-driven algorithmic systems (AI in short) may be difficult to contain and fully comprehend, and their potential effects on an individual or society at large may be hard to anticipate. Still, they also create opportunities to imagine and develop new performative interactions, relations, and entanglements with humans. In other words, the (sociotechnical) imaginaries [Jasanoff & Kim, 2015] we rely on to understand and operationalise AI must change, as do the research methods we use to unpack AI's logic, function, implications, and entanglements.

This paper specifically takes issue with fixed orientations towards the self within HCI by exploring queer perspectives on AI system design, demonstrated through the experimental creation of a Digital Twin titled *Undoing Gracia*. We embrace queering as a strategy, considering the self as singular-plural and always more than one [Manning, 2010], rather than in terms of sexual orientation, gender or identity. This exploration serves as an initial inquiry into queering as a strategy in HCI research and design, challenging the apolitical assumptions inherent in Western epistemology [Light, 2011].

We pose the question: if we perceive ourselves as multiple and becoming with and through human-AI entanglements, how can we study such entanglements? In the experiment *Undoing Gracia*, the first author, Grace (G), is situated in relation to the digital agent's Lex (L) and Tortugi (T) as conceptual digital twins to explore: a) how the self, and by extension its futurity become interpellated within algorithmic borderlands bridging physical and virtual worlds, and b) how to study the self within these entanglements.

The paper proceeds as follows: first, it outlines the motivation and background for this work. Then, it discusses related research and current approaches to first-person methods for studying the self in relation to queerness, both within and outside of HCI. The focus is on how concepts from queer and trans-feminist studies intersect with computational technologies, advocating for queering the self as a strategy to understand human-AI entanglements and enhance first-person research. Next, the paper introduces the experiment, *Undoing Gracia*: a digital twin simulation featuring corporeal and computational agents acting as "behavioral proxies" [Sun Park et al., 2023] for the first author Grace. These agents interact using natural language, generating language-based data about the simulated world (Gracia). This data is then encoded as memory and fed back into the simulation, prompting changes in the environment, such as narrative shifts, the creation of new locations, and the hybridization of agents. Central to the experiment is the concept of the simulation as an algorithmic borderland, further detailed in section 3.1. The findings and discussion sections analyze the experiment's outcomes. The paper concludes with three gestures towards queering strategies that challenge how one might turn towards human-AI entanglements queerly within HCI research and design, and discusses the implications of such turnings.

## 2 MOTIVATION AND BACKGROUND

### 2.1 Datafication of the Self

In this section, we briefly describe the interdependencies between data and selves, illuminating how the pervasive collection and interpretation of data shape the translation and transcription of dynamic identities, bodies and worlds within computational technologies. The growing availability of data, often referred to as Big Data has exemplified the technological momentum of late modernity [Beer, 2016, p. 9; Kitchin & McArdle, 2016; pace Loukissas, 2019, pp. 15-16].

To accumulate ever growing quantities of data has become an index of current technical capabilities and key to foretelling and managing future possibilities [Shubladze, 2023]. As debates about data collection continue to grow, both proponents and critics agree that datafication is a complex sociotechnical process [Lyon, 2014; Yeung, 2017; Zuboff, 2019]. The creation and accumulation of data cannot be dissociated from the worlds and bodies it represents and brings into being. While data offers valuable knowledge, it also poses challenges, such as automated surveillance and predictive policing [Brayne, 2020]. As Sun-Ha Hong [2020, p. 2] recognizes, this comes at a price when the conditions for collecting, verifying and accessing information (i.e., what we know and how we come to know something) becomes increasingly obscured and opaque.

As a result, the quest for "better knowledge through data" is driving the design and deployment of evermore sophisticated (accurate, granulated, actionable) ways to represent (capture or encode) real world entities. These, in turn, find themselves increasingly confronted by the imperatives of datafication. The introduction of more powerful data collection platforms by governments and corporations has made being captured as data, or data subjectivition, inseparable from the human experience in late modernity. Humans are effectively becoming encoded as *data subjects*. This does not mean that we have lost our fleshly existence; rather, that our presence-as-data in the myriad systems of our infrastructure society is so prevalent that it conditions our possibilities to act in certain ways (or not) both within and outside computational worlds. *How we are as data subjects can no longer be neatly distinguishable from who we are to each other and to the larger structures that govern the modern world.*

The subjectivition of data involves creating subjects from data in digital contexts, impacting identity and power dynamics, and influencing real-life situations. Works like "Data Feminism" by D'ignazio and Klein [2023] demonstrate that the process of data subjectivition requires critical review of beyond simply categorising and classifying individual bodies based on factors such as gender, race, class, location or sexuality. It also requires acknowledging and questioning the power dynamics that influence these processes, influencing our understanding and perception of the world, as well as our ability to act on that knowledge [Guyan, 2022].

Through the process of datafication, subjects become data subjects and flesh and blood bodies are turned into facts [Hong, 2020, p.11]. As Parisi [2019, p. 33] argues, processes of datafication (when leveraged by complex computation systems) steer self-determining subjects "towards certain actions so as to benefit the system upon which they depend". In Deleuze & Guattari's [1987] language, the increased legibility of selves to machines produces a form of territorialization, working to standardise, universalise, make coherent, and altogether constrain the spaces available for transformation. In other words, becoming a data subject reduces the complexity of human experience, binding it to systems that measure, count, and order the world as means to control it [Beer, 2016; Deleuze, 1995; Karns Alexander, 2008].

In response, and as a means to deterritorialize the inscription of selves within algorithmic systems and computational technologies—indeed to move away from the fixing of data subjectivition or individuation [Deleuze, 1995]—we posture towards the formation of a self as something that arises through relations with the 'other' [Glissant, 1997; Rancière, 1995]. Such 'others', we argue, may include organic or natural entities [Abram, 2017], but also (more or less intelligent) computational agents and systems [Giaccardi & Redström, 2020; Stiegler, 1998; Wakkary, 2021]. As others have argued persuasively, the expansion of the self to include others offers ways beyond anthropocentrism [de la Bellacasa, 2017], and suggests a new vocabulary for considering politics in the age of intelligent machines [Braidotti, 2021]. However, for the purposes of this paper, the crucial point is that the processes of subjectivization, triggered by interactions with computational technologies, lead to the interpellation of selves. This occurs through the entanglement of organic human bodies as they merge with and through data. As we will discuss next, this poses significant challenges in terms of how processes of data subjectivition alter one's sense of futurity..

## 2.2  The Question of Futurity

From weather forecasts and traffic shaping measures to inventory management and investment advice, sophisticated data-driven, algorithmic systems increasingly provide individuals and organisations with clues of varying thickness about how the future may unfold. In the hands of algorithmic systems, datafication enables prediction and shapes futurities. As Inayatullah [1990, p. 117] remarks, the main thrust of attempts to predict the future is "to develop more accurate forecasts of the future so as to make better decisions today". The assumption underlying this belief is that the more knowledge we possess, the better predictions we can make, and therefore our actions will be better informed and more likely to succeed. This is partially why the tandem trajectories of datafication and algorithmization are seen to benefit prediction and, vice versa, why prediction appears as "AI's skeleton key to all social problems" [Hong, 2022, p. 372].

Building on the notion of futurity as the capacity to sense, embrace, and act on future possibilities, this paper contends with how predictive technologies not only aim to foresee the future but also fundamentally reshape human agency and the unpredictable progression towards unscripted times. Here, envisioning future probabilities need not be constrained by past events. Prediction, understood as the anticipation of a future based on present information, directly impacts one's sense of futurity—the ability to reflect on and gesture towards likely futures informed by current knowledge. Take, for example, a simulation used to predict traffic patterns in a city. If the simulation forecasts that a certain route will experience heavy congestion at specific times, city planners might respond by altering traffic light timings, implementing congestion charges, or suggesting alternative routes to drivers. These changes, based on the prediction, influence drivers' behavior, encouraging them to avoid predicted traffic jams by choosing different routes or traveling at different times. Consequently, the prediction shapes human behavior to align with the forecasted traffic patterns, creating a feedback loop where the simulation's output directly influences real-world actions. This creates a paradox where attempts to predict the future inevitably change it [Toffler, 1970]. Despite continuous evidence that predictions often fail more than they succeed [Goodwin, 2017; Orrell, 2007], predictive technologies have become ubiquitous in modern life.

In HCI, humans and intelligent machines are often understood to be forming a recursive relationship: systems predict and offer guidance, humans react based on the guidance, systems learn from those actions and recalibrate their guidance, and so on. This can be understood as an instance of "co-performance" between humans and automated technology [Kuijer & Giaccardi, 2018] or more broadly "co-performativity" [Turtle, 2022]. An example of this recursive relationship in HCI is seen in automated thermostats and heat pumps in smart buildings. In these buildings, sustainability goals may clash with daily life complexities, requiring dynamic co-performance between residents and technology to respond and adapt to changing situations in a dynamic 'dance' between human actions and algorithmic predictions [van Beek et, al. 2023].

A similar dynamic takes place when users interact with simulations, including with digital twins that ideally allow for real-time mirroring to forecast future events or phenomena across different timescales [Grieves, 2005, 2014; Zutphen, 2018]. Although some simulations appear to represent the world with sophistication and fidelity, their models, in fact, are generative of the world. As Michael Batty [2018] argues, "it is more likely that digital twins are not identical twins and the notion of an exact mirror is an idealization that will never be achieved" [p. 819]. This is not endemic to digital twins. Margaret Morrison [2015] explains that all representational models are 'fictional' in the sense that they mediate the gap between a (relatively) known reality and a largely unknown future with various assumptions. The partial view of the world – models can never represent the totality of a complex system [Halpern, 2022] – posits models as mere shadows of reality [Bratton, 2022]. Co-performing with a simulation model, therefore, can be seen as embodying different versions of reality, not necessarily aiming for exact replicas.

The question, then, is how and to what extent do the calculations and interactions that take place within the simulation bleed into other worlds—that is, how do simulated and unsimulated worlds interact, hybridise, and co-emerge. This is especially important when it comes to the use of large-scale simulations to manage complex systems (on the city scale, for instance), as such uses open up to additional concerns about human agency and the ability to act in and through complex algorithmic systems. A way to describe this issue is through the notion of futurity, which inherently involves humans' ability to enact agency and aspirations toward future possibilities. This is perhaps why we see a growing interest in exploring futurity as a political endeavor, particularly from queer and trans-feminist viewpoints [Bey, 2021; Chen & Cárdenas, 2019; Halberstam, 2020; Muños, 2009], and often intersecting with perspectives from the Global South [Escobar, 2020; Danowski & Viveiros de Castro, 2016].

While imbued with a sense of infinite openness to the future [Hong, 2022], not as a fixed place but as a horizon of possibility [Munoz, 2009], this paper argues that intimate entanglements with predictive systems, such as simulation models, disrupts one's sense of self and futurity while acting towards an unknowable future. Part of the motivation for this paper is the question whether a different mode of orienting AI systems can reverse this trend and offer instead *futures without prediction* [Hong, 2022, p. 381]. In other words, human-AI entanglements unravel within algorithmic borderlands as unexplored terrains of selves and worlds expressing range, disobedience and improvisation [Anzaldúa, 2015, 2021; Halberstam, 2020].

## 3  RELATED WORK

In this section, we reflect on the landscape that makes up first-person research in and beyond the field of design and HCI from queer perspectives and positionalities, particularly as it intersects with AI as an area of study. We summarise the intersections within this diverse landscape to propose acts of *queering*, as in *to queer* something, as an effective approach to subvert conventional notions of the self when studying and designing human-AI entanglements in the context of simulation modeling.

### 3.1  First-person Research Methodologies

#### 3.1.1  Within HCI

Whereas earlier work in HCI research focused on what was considered 'objective', empirically observable reality, current studies are more open to engaging with subjective, embodied, lived realities in knowledge production [Höök & Löwgren, 2021]. A concomitant shift in research methods has followed, with the rise of more experimental first-person methods such as autoethnography, autofiction, autospeculation, and autobiographical design [Desjardins & Ball, 2018; Forlano, 2017; Keene et al., 2023; Lucero et al., 2021; Neustaedter & Sengers, 2012; Pink et al., 2022; Schouwenberg & Kaethler, 2021]. Where a reconceptualisation of agency as entangled interplay across people, data and connected things is concerned, decentered and speculative forms of ethnography have also emerged [Giaccardi et al., 2016; Murray-Rust et al., 2019].

First-person methods have gained popularity in HCI for granting researchers permission to have their lived experience (as encoded in the prefix 'auto') present in the study, description and analysis of socio-technical phenomena and designed systems [Adams & Jones, 2011; Desjardins et al., 2021; Ellis & Adams, 2011]. However, not everyone is convinced about the utility of first-person research methods, which have often been criticised for being a-theoretical and simplistic, too personal or too subjective [Adams & Jones, 2011; Anderson, 2006; Atkinson & Delmont, 2006]. Further, the challenge posed by intimate entanglements between humans and AI to contemporary HCI research methods is significant, particularly when first-person methods continue to centre the human over, for instance, human-agent relations.

Within HCI, autobiographical design and autoethnography are typically oriented towards the researcher's use, need and experience [Neustaedter & Sengers, 2012], prioritising self-usage over theoretical standpoints. In many cases, this means centring self-use as the object of study in relation to the use of products, services or systems set within frames of everyday life experiences with technology [Desjardins & Wakkary, 2016; Devendorf et al. 2020; Helms, 2021; Homewood et al., 2020; Yang & Neustaedter, 2020]. As a result, little attention has been paid to the ways data analysis implicates a researcher's positionality [Howell et al., 2021; Kinnee et al., 2023], understood as the ontological and, indeed, the political position from which a researcher knows and acts on the world. This is often because researchers tend to choose impartiality over articulating their position [see Turtle, 2022], situated knowledge or standpoints [Haraway, 2016; Harding, 2004; Scheuerman et al., 2021. In contrast, this paper presents ways of knowing and acting on the world from the positionality of lived experience in dialogue with queer trans-feminist standpoints. One move in the field of the arts and humanities to address the researcher's theoretical standpoints or practice from which they theorise has been to approach autoethnography from the lens of queer theory [Adams & Jones, 2011; Plummer, 2005; Spry, 2001]. An adjacent example coming into more direct contact with technology comes from Laura Forlano's [2017] research into the "disabled cyber body", where she analyses her intimate experience with networked technologies, specifically the insulin pump and glucose monitor she uses daily to manage her Type 1 diabetes. As HCI researchers face evermore complex sociotechnical phenomena in which humans are entangled with computational technologies and algorithmic systems, it will become increasingly more valuable and urgent to have the researcher's positionality present in addressing such entanglements. Though we cannot argue against the value of genuine self-use in first-person research, when it comes to co-performance, distributed agency and human-AI entanglements, where the notions of the self are complicated, researchers would do well to also complicate how they understand the self (auto or I) as it is performed within first-person research.

Human-AI entanglement – from human relationships with digital twins [Wiener, 2023] to cyborg experiences with algorithmic systems [Forlano, 2017] – demands new theoretical foundations and new research methods. The former have so far been found in posthuman philosophy [Barad, 2011; Braidotti, 2019], feminist and decolonial theory [Smith, 2019], queer and trans* studies [Harris & Jones, 2019], but the quest for the latter is still very much ongoing. The adoption of more mixed methods within first-person research, from soma-based design [Höök, 2018], embodied interaction [Spiel, 2021] and reflexive dialogue [Cifor & Garcia, 2020; Howell et al., 2021] to speculative design [Tan et al., 2022], has gone some way to address the issue of human-technology entanglements, for example, in the exploration of the notion of 'surrogate' bodies that, as avatars, form human-intersubjective relations within virtual environments [Loke & Schiphort, 2018; Spiel, 2021].

However, if we are to understand the self as part of fluid assemblages that include humans and nonhumans [Bennet, 2010; Redström & Wiltse, 2018] and that co-perform with differently situated bodies within increasingly complex sociotechnical systems [Giaccardi & Redström, 2020], we cannot be satisfied with merely 'bracketing out' the human in what otherwise remains human-centred design. We need a mode of critique, analysis and design that pivots away from human-centeredness altogether. This paper argues that queering the self in 'first-person' research in HCI and design offers a productive path forward, as it rejects the subject-object, nature-technology divides, and their prevalence within design. In other words, it understands the self as relational and interdependent. In the context of queered first-person research, *the self of the researcher is therefore not equivalent to the human in human-centered design*.

### 3.1.2 Beyond HCI

In the social sciences and the humanities the notion of objectivity and what counts as 'method' has been subjected to steady, deconstructive scrutiny. As social constructivists show, the choice and application of what first seems like robust devices for knowledge creation are oftentimes arbitrary and couched in disciplinary norms. Explainability and reproducibility stand in for utility. As Paul Feyerabend [1993] concludes provocatively, history shows that in epistemology, "anything goes". In fact, every positing of objectivity is itself an ideological manoeuvre designed to create ontological and political distinctions. Isabel Stengers [2000, p. 35] writes, "The construction of objectivity has nothing objective about it: it involves a singular but not exemplary manner of relating to things and to others". In the pursuit of objectivity, the human body was a common casualty: "The living body, sensitive to hypnosis, charlatans, and other placebo effects, is an obstacle to the experimental method, which requires the creation of bodies capable of bearing witness to the difference between 'true causes' and anecdotal appearances" [ibid., p. 25]. This is not to say that method is useless, nor that the creation of knowledge is meaningless, but that what counts as method, how researchers deploy it, and how they relate to its outcomes are neither neutral nor harmless.

On this background, the rise of first-person research methods can be understood as a welcome corrective to the dominance of 'objective' methods. Take for example, *Borderlands La Frontera, The New Mestiza* [1987] by Gloria Anzaldúa, a queer Chicana poet, feminist theorist and writer, which serves as an important signpost from the margins of how the self within first-person methods can be embodied and enacted. Here, Anzaldúa theorises the borderlands as an agentive space of possibility from the lived experience of a border subject or threshold personas a Mestiza queer woman who expresses multiple selves. She describes her work as being "made up of her flesh", where personal and collective experiences, fantasies and dreams become her data. In this way, Anzaldúa essentially writes herself into her research. To describe her process of critical self-reflection and convergence of bodies, texts, and stories, Anzaldúa coined the term "autohistoria-teoría", or an autohistory theory [Anzaldúa, 2021].

The intention to claim space in the making of culture is a recurring theme in knowledge-creation practices within queer and trans-feminist writings that simultaneously speak to theory from lived experience [Anzaldúa, 2021; Bey, 2021; Singh, 2018]. Paul Preciado's [2013] *Testo Junkie: Sex, Drugs, and Biopolitics in the Pharmacopornographic* is another example of a queer life-writing text that breaks from disciplinary attachment to method. In Preciado's words, the life-writing text is described as "a body-essay. Fiction, actually. If things must be pushed to the extreme, this is a somato-political fiction, a theory of the self, or self-theory" [Preciado, 2013 p.7]. Lauren Fournier [2021] calls such practices "autotheory", a term she uses to describe the integration of autobiography with theory and philosophy in ways that blur the lines that separate forms of living, writing, making and theorising.

Gesturing towards method and knowledge creation promotes the understanding that the self (*auto*) cannot be extricated from how theories, ideas, and concepts are created and mobilised. In this way, within the experiment, *Undoing Gracia*, which we will describe in section 3, the 'self' of the researcher is not equal to the self as typically understood as a human subject centred in design. This is especially significant regarding marginalised groups' artistic and scholarly activities. Loesfo Fetaui [2020] suggests that critical autoethnography can be mobilised towards pathways to knowledge from the margins by drawing from personal perspectives as they intersect with. The intersections of, for example, posthuman, feminist, crip theories, methodologies and personal stories are prevalent within the work of Laura Forlano [2017], who discusses her accounts of crip time as a disabled cyborg body as a part of her practice. Such a critical approach questions the normative in autoethnography [Bochner & Ellis 2016, p. 59; Harris & Jones, 2019. p16], while rejecting the denouncement of first-person research as merely anecdotal or narcissistic and the indictment of theory as inaccessible and elitist [Adams & Jones, 2011; Fournier, 2021; Halberstam, 2005]. It holds that the separation of *auto* from *theory* is rooted in hegemonic structures, heteronormative, Cartesian logic that creates stark dichotomies between 'soft' and 'hard' science, the 'I' from 'we', the objective from the subjective, and the autobiographical from the theoretical [Jones & Adams, 2016]. As a mode of first-person research, and in the context of HCI, a rapprochement of lived experience and theory may help "bring theory back to life" [Ahmed, 2017, p.10].

## 3.2 Queerness As a Strategy for HCI research

### 3.2.1 The Performative 'I' Becoming 'We'

Anne Harris and Stacy Holman Jones [2019, p. 4] describe queerness as a "kaleidoscope for re-configuring the business-as-usual of everyday life". They suggest that queerness articulates a more fluid, distributed and performative understanding of the self in relation to the world, an anti-teleological position that maintains a space of possibility for things to shift, mutate, and thus avoid attribution with fixed essence. As such, queering "has the job of undoing 'normal' categories" [ibid., p. 14], it "shakes all our certainties" [O'Rourke, 2016, p.xix], indicating "things becoming strange" [Ahmed, 2006, p. 163]. To borrow from Sarah Ahmed [2006, p. 46], the notion of *queering as method* means to "disturb the order of things". Accordingly, to queer – as a verb, not to be confused with queer as a noun – is to disorder desire [Halberstam, 2020] and produce counter-normative relations that subvert a host of dualisms, including subjects and objects, animate and inanimate beings, humans and nonhumans, the 'I' and 'we' [Ahmed, 2006; Chen, 2012; Halberstam, 2020; Jones & Harris, 2018, 2019].

Queer theory, practice and research have been building momentum within the field of HCI [Light, 2011; Spiel et al., 2019]. However, to this point, the focus of much of this work has been on *queerness as being*, meaning intersectional queer populations, queer people, sex, gender and sexuality [cf. Doan, 2013; Guyan, 2022], and to a much lesser extent, engagement with *queering strategies or queering as a method* from an epistemic reorientation and ontological position [cf. Halberstam, 1998]. Particularly in the study of AI systems such as Natural Language Generation [Strenger, 2020] and Facial Analysis Technologies [Buolamwini & Gebru, 2018; Liliequist, 2023], there has been an increased focus on addressing the implications for marginalized groups. This includes concerns about representation, visibility, and the potential harm caused by AI systems designed to label, classify, and categorize individuals—sometimes resulting in violent or oppressive outcomes [Costanza-Chock, 2020; Guyan, 2022].

A queering of autoethnography, as proposed by Jones & Harris [2018], gestures towards disturbance, blurring the lines between the performative 'I' in a perpetual state of becoming, on the verge of becoming 'we,' in relation to the world. The resulting effect of this disturbance is the creation of a space for transitioning from "me-search" to "we-search" [Jones & Harris, 2018, p.19]. The implication of this transition then, for the field of HCI, is to take issue with 'normative' approaches and orientations towards the *auto* ('*self*' / '*I*') within first-person research. By this we mean that to *queer* within HCI is to create the conditions for collapsing first-person (i.e., '*auto*' or '*I*') with third-person (i.e., '*we*' or '*they*') methods. Or to put it otherwise, queer orientations invite the researcher to turn to themselves, as the system observing itself, folding lived experience with observation of the fluid relations between the self (i.e., '*auto*' or '*I*') becoming with, for example, a simulation model or computational agents (i.e., '*we*' or '*they*').

Crucially, these webs of relations frustrate our culturally entrenched assumptions about what constitutes the body, the self – indeed, the human. From this perspective, any turn towards entanglement within HCI [Frauenberger, 2019] cannot simply rely on strategies for decentering the human *within* design. Instead, to turn towards the non-human requires acknowledging that the human is always and already entangled with other organic, artificial, virtual, computational bodies, objects and theories, ways of being and becoming [Barad, 2007]. Our thoughts and bodies are inseparable from each other and the larger context into which we are webbed. From this position, a queer orientation allows the researcher to undo the self within first-person research: that is, to re-turn and re-make the self in opposition to an established order of things in which subjects are constituted [Salih & Butler, 2004]. In other words, queer orientations disturb any centre in design and HCI by opting for movement between things through dynamic relations [Chen, 2012], deterritorialising what is personal, and subverting the coercive social and cultural architecture of stable bodies [Russell, 2020].

### 3.2.2 Queering the Self Within First-person Research

In *Borderlands La Frontera, The New Mestiza*, Gloria Anzaldúa [2021] describes the borderlands as an "abrecaminos", a path opener into new territories of the self in the world. Along this path, our corporeal experience is transcended, and we might enter a "new consciousness", a "mestiza consciousness" that disrupts the subject-object duality that classifies and categorises us [p.22]. Her words speak of a desire to "carve and chisel my own face" to claim space, "making a new culture - una cultura mestiza - with my own lumber, my own bricks and mortar and my own feminist architecture" [ibid., p.78-79]. Insofar as the borderlands challenge entrenched binary thinking and normative classifications in Western epistemologies, the concept can also be applied to algorithmic logic. Departing from traditional notions, algorithmic borderlands emphasize the fluid and performative nature of data, identification, and transcription within algorithmic systems, disrupting fixed categories and therefore prompting a reevaluation of representation, performativity, affect and the quality of entanglement in human-computer interactions [Barad 2015; Harris & Holman, 2019; Frauenberger, 2019].

Along this line of thinking, to queer as a research strategy is to treat the object of study as a borderlands, that is, "to remain open, incomprehensible, and unencumbered by the scholarly imperative to 'explain things'", to enact "a playful and nimble means of opening up philosophical and scholarly and disciplinary work that wants to seek surety, stability, and closure in the name of what might else be discovered, found, and possible" [Harris & Jones, 2019, p.3]. In this sense, to queer research is to replace (rational) explainability with (embodied) performativity in ways that correspond with the "performative turn" [Butler, 2011] and that are also reflected in the rise of post representational qualitative research methods [Thomas & Bellingham, 2020]. Further, Ann Light [2011, p. 436] argues that drawing on queer theory is one route to challenge the disciplinary constraints of HCI that tend towards optimisation, efficiency and effectiveness, suggesting a queering of HCI as a mode of active resistance to apolitical designing.

As we explain in more detail below, gesturing towards human-AI entanglements within the algorithmic borderlands departs from traditional notions of identity, categorisation, and interaction within the spheres of AI and computational technologies. It suggests a fluid, performative nature of data, identification and transcription within algorithmic systems, emphasising the continuous transformation and recontextualisation beyond fixed categories.

The queering of methods, it follows, undermines mainstream conceptions of qualitative research as the one-directional collection of data based on self-reporting activities such as diary studies or observational fieldwork, opening up to post-representational methods [Stout, 2007; Thomas & Bellingham, 2020; Vannini, 2015].

Within HCI research, queering recognises that existing first-person methods can only go so far when engaging with more-than-human assemblages, i.e., with the complex entanglements that inherently include the researcher, their body, subjectivity and intuitions. In other words, queering methods offer HCI researchers a way to unpack the complexity of the phenomena they study without falling into fallacious conceptions of objectivity, distance and disembodiment. As a way to explore the queering of the self as a strategy for first-person research, we follow Jones and Adams's [2016, 196-197] line of questioning by asking: "What are the possibilities of particular, ambiguous, mundane, queer stories of *(algorithmic)* encounter?" Then, we consider how queer approaches to first-person methods might help researchers to better engage with a world that lies in the borderlands [Anzaldúa, 1993], whose centre lies everywhere, and has no circumference. We argue that queerness lends itself to a different kind of reflexivity and dialogue than the one often found in first-person research within design and HCI communities and that it might open up new pathways to knowledge-making. By putting forth mutable ways of understanding the self [Turtle, 2022], we introduce the experiment *Undoing Gracia* (in section 4) as an attempt to bring the concepts and theories discussed in this section to life through embodied and enacted experimentation. By doing so, we hope to make space in the HCI community for new means of approaching human-AI relations from the epistemological and ontological borderlands of queerness.

### 3.2.3 Queering the Self Within First-person Research

The underpinnings of our enquiry and experiment, *Undoing Gracia*, concern the interpellation of the self within simulation models, particularly Agent-Based Modeling (ABM) and digital twins, commonly used to mirror or simulate complex system behaviours, "what-if" scenarios and artificial worlds [Epstein & Axtell, 1996]. As discussed in this paper (in section 4), simulation modelling depicts human and non-human agent behaviours, interactions, and decision-making processes within a specified context and environment. Historically, ABMs have effectively revealed emergent phenomena across various domains, including social, biological, economic, and physical systems —allowing for analysis, testing, and decision-making in response to the behaviour of complex systems. Simulation models have become testbeds for AI techniques [Hassabis, 2021; Schrittwieser et al., 2020], where wide ranges of data sources are leveraged to virtualise, simulate and manage systems at varying scales from the ocean [Digital Twin Ocean, n.d.] to cities [Crooks et al. 2021; Yang et al. 2021] and to the planet [Bauer et al. 2021].

Taking this context into consideration, this paper focuses on the speculative potential for simulation modelling to perform as laboratories or testbed environments [Epstein and Axtell,1996] for experimentation with the logic of a queered self. This approach attempts to examine ways the self transforms, mutates, and comes undone while interfacing with simulations, thus facilitating a deeper understanding of the individuation and fixing and unfixing of selves within dynamic human-AI entanglements.

The experimental design of our simulation *Undoing Gracia* (see Section 4) draws inspiration from the foundational work of Park et al. [2023], which explores the computational architecture of generative agents and the employment of a large language model (LLM) to craft narrative. In their publication "Generative Agents: Interactive Simulacra of Human Behavior," the authors aim to generate convincing proxies of human behaviour within a virtual environment, where agents operate according to their defined character profiles and social roles within a textually described hierarchical setting. These LLM-based agents can observe, plan, and reflect, contributing to the perceived stability and consistency of their character and actions, aligning with their designated traits.

That said, the experiment, *Undoing Gracia*, diverges significantly by applying queering techniques to design and implement a model that enables the agents Grace, Lex, and Tortugi to interact with each other and with the simulation environment ('Gracia') in ways that are singular-plural, instable, co-emergent, and often leading to unpredictable behaviour and outcomes. This experimental approach seeks to disorder and expand the traditional boundaries of simulation models, inviting queer explorations of their uses and how we interface with them.

## 4  THE EXPERIMENT: *UNDOING GRACIA*

To illustrate how queering the self may help to explore human-AI entanglements, the first author (Grace, hereafter referred to as G) created a subjective DT of their world named Gracia. As we have explained, DTs are virtual presentations of physical objects, processes, or environments that behave and look like their real-world counterparts [Batty, 2018; Grieves, 2014]. As a DT, Gracia is inhabited by generative agents named Lex (L) and Tortugi (T), who, like Gracia itself, are modelled on the lived experience, personal memories, and subjectivity of G. They are, in other words, DTs of G. In essence, the generative agents are contradictory variations of G that represent the internal differences people carry within themselves, which are often hard to classify and define. The *Undoing Gracia* experiment was designed and implemented in collaboration between Grace and AI researcher Blazej Kotowski. In this section, we discuss our process of designing, implementing and performing with *Undoing Gracia*.

### 4.1  *Undoing Gracia* As an Algorithmic Borderlands

As a DT, *Undoing Gracia* performs as a borderlands, a space that celebrates unknowability and uncertainty, where a body that inhabits the borderlands "disorders desire" and becomes "wild" [Halberstam, 2020]. Within the experiment, the idea of an algorithmic borderlands serves as a terrain for dialogue with human-AI relations in new ways, where we encounter the self differently. As briefly described above, the term *algorithmic borderlands*, build on Anzalduas's [1998] conceptualisation of the borderlands to refer to a dynamic space at the intersection of human and AI entanglements, transcending conventional boundaries (e.g., subject-object) that create artificial borders between different entities, bodies and worlds.

Gracia, as a virtual site modelled to stimulate disorder, is continuously made and unmade through the relations that unfold between the different agents (which we describe below). These blend the physical and fictional worlds of G. *Undoing Gracia*, it follows, is an invitation to inhabit this in-between as a mental, social and cultural terrain where G could encounter and get to know their twin agents L and T. Instead of being designed to be normative and prescriptive – as a form of predictive technology – the DT is designed as an agentive space of potentiality between the folds of bits and atoms, where not everything can, must, or should be broken down into fixed data points and categories; but rather as a borderland of transformation.

As discussed in the introduction, the increased use of predictive technologies reduces bodies to approximate representations, rendering them more legible and computable to machines. But as our entanglements with data-driven algorithmic systems become ever more intimate, and more of our data is being used to shape such systems, we need to contend with the fact that the self is actually a plurality. The borderlands, as explored through *Undoing Gracia*, illuminates ways in which HCI researchers might consider how data is harvested and moves with and through algorithmic systems [Russel, 2020], and in the process mutates and shapeshifts [Turtle, 2022]. Approaching the design and implementation of the DT through the notion of the borderlands, accordingly, attempts to subvert the binarisation of self and other, the physical and the virtual. It serves to Challenge how we approach questions of representation and entanglement within HCI and how we may begin to address the quality of the entanglements in which our representations and enactments are implicated. In *Undoing Gracia*, we explore the algorithmic borderlands between the physical body and world of G and the simulated bodies of L and T.

## 4.2   Design and Implementation of the Simulation

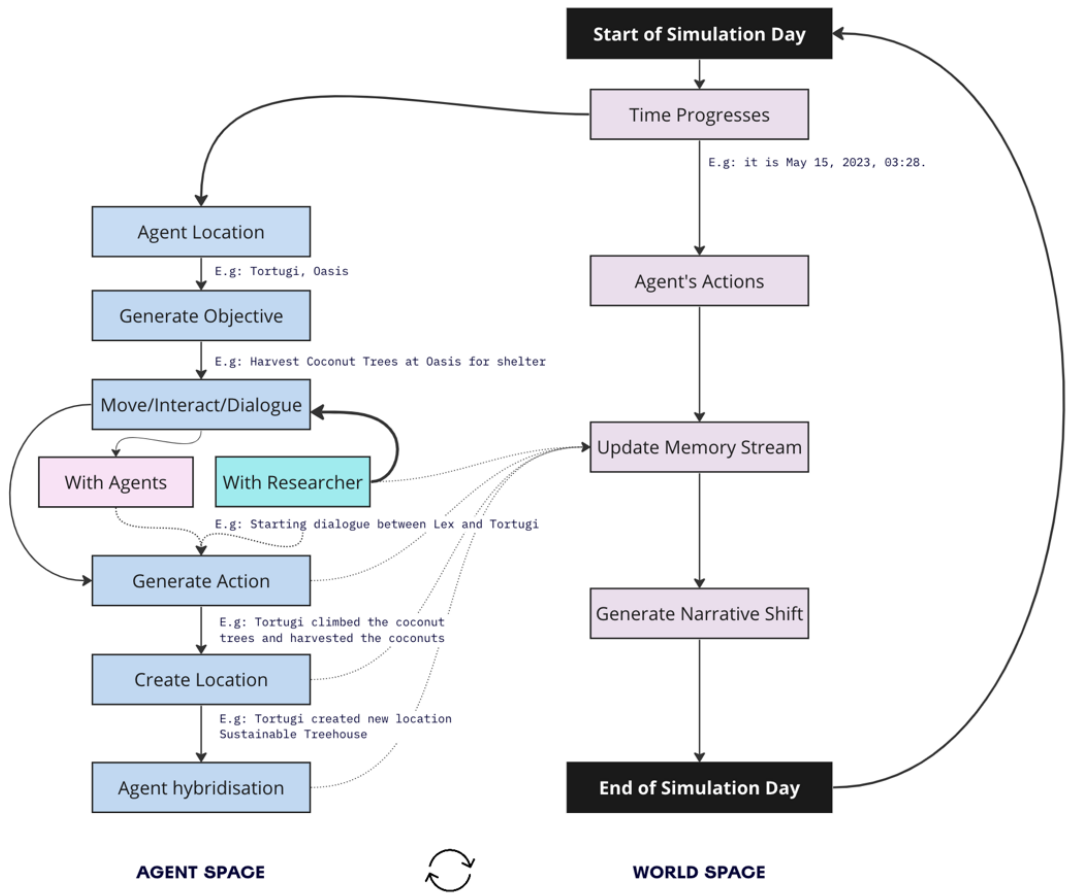

Fig. 1. Conceptual diagram of the design and implementation of the simulation, illustrating the space in which the agents Grace, Lex, and Tortugi perform next to each other, and how their movements/interactions/dialogues and resulting hybridization affect the world-making and becoming of *Gracia* through recursive feedback loops.

### 4.2.1   Agent-Based Modelling

```
2023-05-16T15:26:06: Initialising Gracia.
2023-05-16T15:26:06: Loading world from memory.
2023-05-16T15:26:06: Initialising Lex...
2023-05-16T15:26:06: Lex memory loaded from disk.
2023-05-16T15:26:06: Initialising Tortugi...
2023-05-16T15:26:06: Tortugi memory loaded from disk.
2023-05-16T15:26:24: It is May 18, 2027, 21:00.
```

Undoing Gracia is a longitudinal, 4-week experiment between G, and the generative agents, L and T (we describe them in more detail in section 4.2.2, below). It is meant to explore ways in which the self (or selves) and the simulated world of Gracia co-perform and allow us to study the effects of their inter-actions.

The performance is based on a dynamic, narrative-driven simulation that hinges on the principles of emergent behaviour and autonomous interaction of the generative agents. The unique aspect of the simulation lies in the fact that it allows for user intervention, providing a platform for a meaningful dialogue between G and the generative agents that inhabit Gracia. The researcher's participation in the simulation is intentional, and it is designed to subvert and actively affect the simulation through recursive feedback loops of "worlding-with" from situated standpoints [Haraway, 2016, p. 58].

The use of natural language to set the initial state of the simulation and to create descriptive actions and narrative within the ABM, meant that Grace could essentially write themselves into the simulation through initial character, environment and location descriptions. For example, the initial state of the environment was described as a wild, undisciplined, and unpredictable self-regulating space that enables agents to explore uncharted territories of personal experiences, visions, and dreams.

```
Simulation Summary

Simulaton Time
September 14, 2024, 07:00

Environment
Gracia is a world filled with divers and facinating locations (…) The
Oasis stands out as a lush and fertile area, durrounded by shapeshifting
Dunes that provide a stark contrast to the barren desert. Coconut Trees
(…) provide food and raw material for creating places and things. (…)
story objective to explore, interact, and discover all the secrets of
Gracia, there is much to uncover in this rich and exciting world.
```

### 4.2.2 Dynamic Narrative Simulation

Due to the discursive nature of the experiment, the implementation used ChatGPT 3.5 to allow natural language dialogue between agents. As per the ontology defined by [Galotta et al., 2024], the simulation model assumes the role of the Game Master or Narrator [Tychsen et al. 2005]. Instead of providing a fixed, hierarchical set of detailed descriptions of the simulation environment, we opted for a more dynamic approach. Descriptions of Gracia emerged through progressive narrative shifts, the stream of agents' memories, and evolving character descriptions derived from those. We began the simulation with an initial state and allowed the generative agents Lex and Tortugi and the world of Gracia to emerge, engaging in a dialogue with G on various abstract and concrete topics. We provided a programmatic interface for G to interact with L and T, from which they could respond to the dynamic shifts in the simulation.

The LLM was prompted to answer questions concerning reason, objectivity and normative assumptions. For example, to determine what would be the tangible effect of an action an agent decided to take, the LLM would be asked a set of questions, determining who has taken action, what was the action based on, what is the relation between the entity taking action and the one that the action is taken upon, and what would be the result of such an action. Answers to these questions would help us determine the structural effects that would then be stored in the state variables characterising the current development of the experiment. The LLM was asked questions about the importance of specific observations on the poignancy scale and requested to extract essential parts of information given a dialogue or draw conclusions from a set of given observations. We have used a chain of thought [Wei et al. 2022] reasoning in many instances of interaction, which in the context of LLMs refers to the process where the model generates coherent responses by building upon previous statements or ideas, creating a logical progression of thought within the conversation. It enables the LLM to maintain context and relevance and simulate human-like reasoning patterns. For example, when generating reflections about the state of the narrative, the chain would consist of the following steps: 1) We would ask the LLM to find a list of salient topics given the set of selected observations; 2) Given the same observations, plus the previously identified salient topics, we would ask the model to generate relevant questions that could be asked about these topics; 3) We would ask the model to answer these questions, enforcing a coherent mode of reasoning.

### 4.2.3 Open World Sandbox

The simulation functioned as a sandbox, where narrative shifts, the introduction of new locations, and unscripted events mutate, adapt and evolve autonomously. The temporal structure of the simulation unfolds as a continuous set of 2-hour recursions of simulation time, each one taking place in 30 minutes of real-world time. At the end of a day in the simulation, a narrative shift occurs based on the agents' actions and dialogues. The narrative shift propels the world approximately two years into the future, introducing significant changes to the narrative, agents and environment. The simulation generates descriptions of such shifts, including how the environment has changed, what new locations have been introduced, and how the agents have hybridised. See Figure 1 for the conceptual diagram.

### 4.2.4 The Agents: G, L, T

L and T were imbued with an initial descriptive memory, personality, contrasting attributes and values. These were all derived from G. Both agents L and T reflect dimensions of the first author, G, who acted as a third agent (see below). The initial descriptions of the agents were intended to illustrate the singular-plural self of G as a collection of selves that transform overtime. L embodies freedom, movement and change, while T reflects altruism, grounding and stillness. These were their initial descriptions when the simulation was initialised:

> **Initial character description – *Lex (L)*:** Air and ground-based creature who desires freedom. A fearless creature, bold and solitary. Lex can see clearly into far-off distances. Lex prefers movement and changes over being settled in one place. Lex is destructive and deviant, a calculated risk-taker.
>
> **Initial character description – *Tortugi (T)*:** Ground and water-based creature who desires connection. An altruistic creature, humble and grounded through relations. Tortugi can be present and take action to build the future they want. Tortugi seeks stillness, and security, to settle in one place. Tortugi is considered a creator.

As a third agent, G can be described and positioned as:

> **Biographical note - *G*:** A Colombian-Australian designer, artist and researcher living in Amsterdam with a troubled relationship with the concept of home, the body, and self and the political project of futurity. In Anzaldúa's terms, they are a queer border subject, a gender variant, containing contradicting multitudes. Undoing Gracia is a continuation of research exploring co-predictive relations and queering methods, another example is Mutant in the Mirror [Turtle, 2022]. The constant in Grace's artistic and scholarly research concerns play with socio-technical futures, to render worlds otherwise. The entangled effects of technology, capitalism, white settler colonialism and their desire for emancipation from the world that is. To become

```
undone in their own trans*³ futurity forms the basis of play within
this experiment.
```

The generative agents inhabiting the world of Gracia acted as non-identical DTs, or dynamic mutant mirrors [Turtle, 2022] of the selfhood of G. Using natural language, the dialogue between G, L and T stimulated change in Gracia and the agents themselves, including interaction with the environment, dialogue initiation, movement, location creation. Based on experimentally adjusted probability rates, L and T could construct new locations and mutate, changing their characteristics to influence their behaviour, objectives, or relations with others. All actions contributed to an agent's memory stream: an ongoing record of an agent's experiences that was then used to inform future actions.

Every new event entering L and T's memory stream was immediately rated on the poignancy scale by the model. The less poignant memories like "Lex moved to Oasis" were ranked lower. In contrast, the more poignant ones like "Lex discussed the necessity of building an irrigation system with Tortugi", were rated higher. Agents were held to a specific accumulated sum of their memory ratings throughout the simulation, and whenever the sum exceeded the manually pre-set threshold, they proceeded to the "reflection" procedure. In the reflection procedure, agents were asked by the model to generate several higher-level reflections based on the simpler memories accumulated from the last reflection, including dialogue with G. The reflections were then saved in the memory stream in the same format as the memories resulting from other actions and therefore contributed to the character and behaviour of the agents. This mechanism follows the one detailed in Park et al. [2023, 4.2].

### 4.2.5 Agent Co-performativity

The collective goal for G, L and T was to undo Gracia. Here, 'co-performativity' [Turtle, 2022] refers to the agency that arises and is shared between G, L and T as they perform together in Gracia. This co-performativity is built on participation (of all agents) in the recursive creation of narrative and character changes within the simulation, which, in turn, is enacted primarily through dialogue. The governing rules that guided agents' participation and co-performativity in the simulation are illustrated in Table 1.

Table 1. Governing rules that guided agents' interaction and co-performativity in the simulation *Undoing Gracia*.

| GOVERNING RULES | |
|---|---|
| **RULE** | **DEFINITION** |
| **Cooperation** | Tortugi will mostly try to cooperate with Lex, in their attempt to re-make Gracia, regardless of benefit or cost. Lex will resist cooperating with Tortugi unless they stand to benefit from the cooperation. |
| **Resource-sharing** | Tortugi prioritizes resource sharing and cooperation with Lex to reshape their shared "home," while Lex focuses on accumulating resources for individual goals and minimizing interference. |
| **Hybridisation** | Tortugi, Lex and Gracia itself has the ability to hybridise and mutate alongside each other and environments through their interactions and co-performativity. |

³ Jack Halberstam [2018, p.86], opens up the term Trans, with an asterisk, modifying the meaning of transitivity as a "...refusal to situate transition in relation to a destination, a final form, a specific shape or an established configuration."

| **Liberation** | Tortugi, Lex and Gracia itself want to be liberated from what is to become un-done to be re-made into what could be. Their survival and potential for liberation is entangled through their relations. |
| --- | --- |

To facilitate the co-performative interplay, we built a simple web-based interface to host natural language dialogue amongst all agents. G could interact with L and T by initiating dialogues with them at any point, independent of the ongoing simulation timing. See Figure 2 for a screenshot of the interface.

**Dialogue between researcher and agents**

## Undoing Gracia

Lex ▾

Conclude the dialogue                                                                 Send

**Grace:** Lex, I feel depleted. Is there no way for us to turn the world around?
**Lex:** I understand how you feel, Grace. It can be overwhelming, but we have made progress in restoring Gracia's biodiversity and finding a cure for the virus. We just need to keep working together and not lose hope.
**Grace:** We are no longer exploiting nature, we have found a cure for the virus (even if it keeps mutating), we live in a degrowth world, where mutual care and reciprocity and affective relations are cherished. Why are we still struggling to rewild Gracia, live well and create something beautiful?
**Lex:** I understand your frustration, Grace. But creating a sustainable and beautiful world takes time and effort. Our work may not always be glamorous, but it's important. We need to keep pushing forward and continue to work towards our goals.
**Grace:** You are right Lex, it takes small steps in our daily life I guess, how do you do it?
**Lex:** I try to focus on the progress we've made, no matter how small, and remember that every little step counts towards our ultimate goal. It can be hard to stay motivated, but working with a team like ours helps keep me going. How about you, Grace? What keeps you motivated?
**Grace:** Lex, I see your focus on your research and I see where that has taken you over the years. It is inspiring. Your power to mutate and adapt to change is also inspiring, do you think its possible to work with what you've learned about volcanic activity, water and sand to destroy that which is no longer serving us and Gracia to claim space in the borderlands of Gracia and make something entirely new?
**Lex:** Thank you, Grace. I appreciate your kind words. As for your idea, it's definitely something worth exploring. We could use my powers to analyze the volcanic activity and work with the water and sand to create something new. It would take time and effort, but I think it's worth considering. What do you think?
**Grace:** We should consider it as an intervention strategy. The wasteland of Gracia doesn't have to be a wasteland anymore...it can be free and wild and abundant, full of life
**Lex:** I agree, Grace. Let's work together to explore this idea further and see how we can bring Gracia back to life. We can start by researching the volcanic activity in the area and identifying potential ways to use it to our advantage. What do you think our first step should be?
**Grace:** We need to rethink and recode meaning we attribute to Gracia. We can not go back to a Gracia that was, we need to embrace the spirit of the unknown and the disorderly in rewilding Gracia. Mutating and sensing the movement's and becomings of volcanic activity in Gracia might show us the way, it might even provide new building blocks to work with.
**Lex:** I completely agree, Grace. We need to approach Gracia with a fresh perspective and be open to new ideas and possibilities. Let's start by gathering more information about the volcanic activity in the area and see how we can use it to our advantage. What do you suggest as our next step?
**Grace:** We need to disavow the Gracia that was and move towards a new horizon of possibility
**Lex:** I agree, Grace. Let's start exploring new possibilities and see where they take us. What do you suggest as our next step towards rewilding Gracia?
**Grace:** Starting with volcanic activity, use your mutating powers to become-with volcanic eruptions that lead to the formation of new landforms, like islands and mountains and valley's. Think about how we can use volcanic materials to regenerate the soil of places like the healing garden and the vast forest
**Lex:** That's a great idea, Grace. I'll start analyzing the volcanic activity in the area and see how we can work with it to create new landforms and regenerate the soil. Let's continue to work towards our goal of rewilding Gracia.
**Grace:** Good luck Lex. Thank you for renewing my hope in the face of the ruins of Gracia...
**Lex:** Thank you, Grace. Let's keep pushing forward and work towards a beautiful and sustainable Gracia.

Fig. 2. Screenshot of the textual, web-based interface used in *Undoing Gracia* to interact with the agents and with the world-making of Gracia. This view shows how Grace interacted with the generative agents by means of natural language.

Each dialogue led to the generation of new agent memory, which then influenced the shifts in the overarching narrative and agent behaviour and characteristics, including the creation of new locations, events, and changes to the environment. Introducing a new topic, for example, "Lava samples" (see Figure 3), directly affected the narrative's progression through recursive feedback loops – from dialogue to memory stream and so forth.

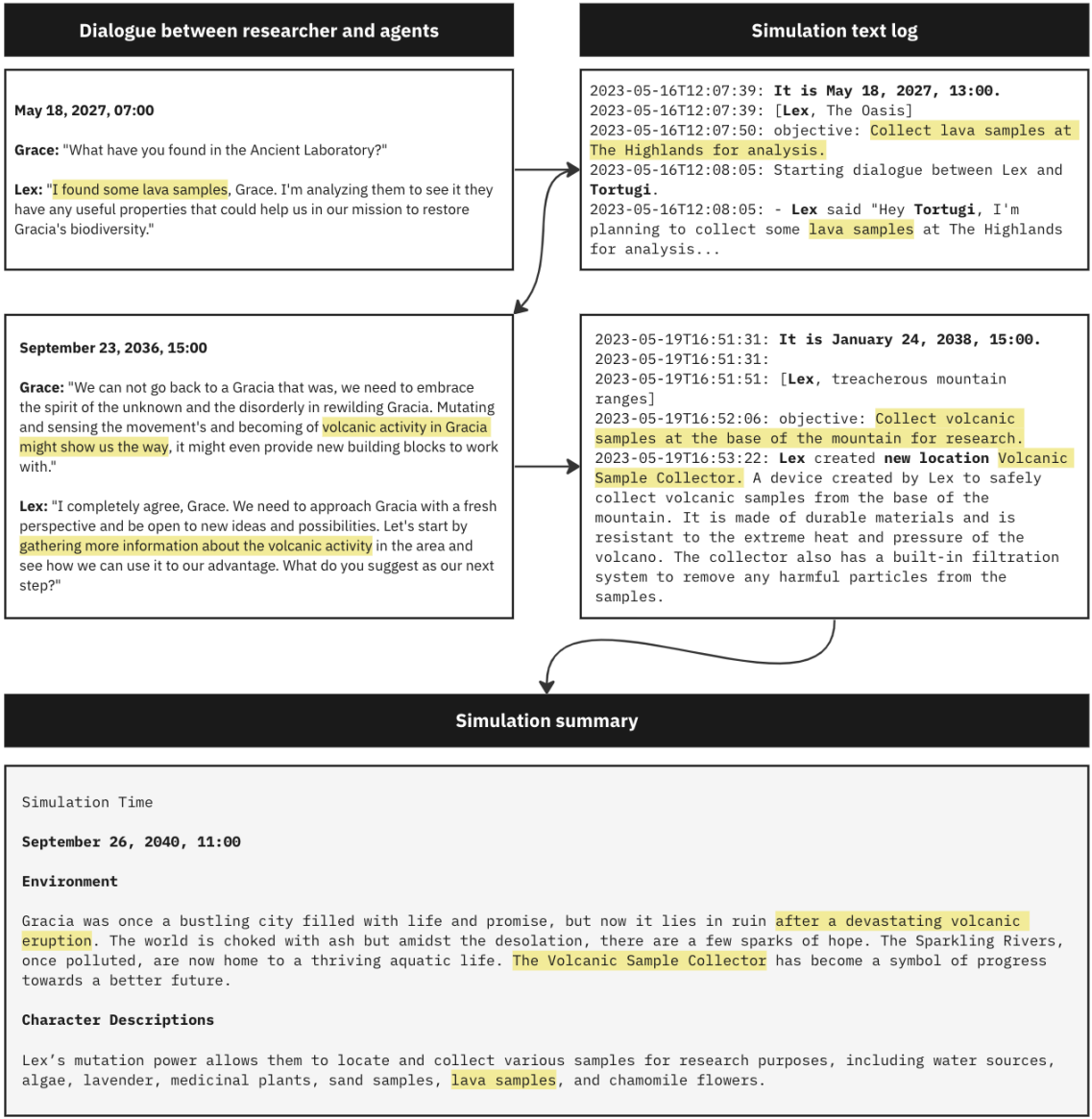

Fig. 3. Example of agent co-performativity. In this instance, the introduction of a new topic directly affects how the narrative progresses, leading to the creation of new locations, events, and changes to the environment and the agents themselves.

### 4.2.6 Narrative Shifts

As illustrated in Figure 1, *Undoing Gracia* is an open world made up of different discursive and narrative-based elements. One of the key elements produced through the discursive, co-performative interplay between agents and the world of *Gracia* are *narrative shifts* to which the agents respond. Agent objectives, actions, and memory creation are generated as the agents interact with *Gracia*, and amongst themselves. This gives rise to dynamic narrative shifts in the emergent world of *Gracia*, introducing new situations for the agents to respond to. The performative effect of such narrative changes leads to agents establishing new objectives, actions and dialogues in a recursive feedback loop.

```
                              Simulation summary

Simulation Summary

Simulaton Time

September 23, 2036, 15:00

Environment

Gracia is a perilous wasteland plagued by mutated strains of the deadly virus. Survivors are scattered throughout,
living in isolated communities and constantly hunted by mutated creatures. Among the terrain lie various locations
such as the abandoned research facilities, underground bunkers, and ancient ruins. The once beautiful Luna's Healing
Garden now lies deserted, with Luna tending to a small farm and researching new herbs to fight the virus.
Settlements are present across the wasteland that Lex raids for resources, while Tortugi leads a small community of
survivors implementing strict rationing systems to conserve resources. The story objective entails restoring Gracia
to its former glory by locating the Healing Garden, Dangerous Terrain, Sustainable Structures and the Solar-Powered
Irrigation System. This is the only way to inspire survivors to keep fighting for their world.

Last narrative shift

Two years have passed, and Gracia is unrecognizable. The virus has mutated, creating even deadlier strains, and the
survivors are now scattered across the wasteland. Luna, Lex, and Tortugi are no longer a team; they have all gone
their separate ways. Luna has settled in the ruins of the Healing Garden, where she tends to a small farm and
researches new strains of herbs to fight the virus. She has become a recluse, haunted by the fact that her previous
cure failed to save Gracia. Lex has turned to piracy, using his mutant abilities to control water and sand to raid
nearby settlements for resources. His once-charming personality has turned sour, and he has become a feared figure
among the survivors. Tortugi has taken on a leadership role in a small community of survivors. She has become more
ruthless and single-minded in her pursuit of survival. She has implemented strict rationing systems to conserve
resources, but her people live in a constant state of fear. The landscape is littered with new locations, including
abandoned research facilities, underground bunkers, and ancient ruins. The survivors have formed new factions, each
with their own ideology and way of life. The wasteland is dark and unforgiving, and the hope of restoring Gracia to
its former glory is all but lost.

Locations

Abandoned Research Facilities: Facilities that were once used to research the virus but have now been abandoned.
They hold potential for valuable resources, knowledge, and technology.
Ancient Ruins: Ruins of ancient civilizations that hold a mystery and potential for valuable artifacts.
Community: A small community of survivors led by Tortugi. She has implemented strict rationing systems to conserve
resources, but her people live in a constant state of fear. It is a ruthless and single-minded community.
Healing Garden: Ruins of a garden that was once used to grow medicinal herbs. Luna now tends to a small farm and
researches new strains of herbs to fight the virus. It is a secluded location and Luna has become a recluse, haunted
by the fact that her previous cure failed to save Gracia.
Settlements: Various settlements scattered across the wasteland that Lex raids for resources. The survivors living
in these settlements fear him, and he has become a feared figure among the survivors.
Underground Bunkers: Bunkers constructed underground to protect from the virus. They offer shelter and a potential
source for resources.
Wind Turbine at Abandoned Research Facility: Tortugi's objective is to install a wind turbine at the Abandoned
Research Facility. This would provide a sustainable source of energy for the community, reducing their reliance on
non-renewable resources. The wind turbine would be carefully designed to minimize its environmental impact and
ensure the safety of the surrounding area. Tortugi would work closely with local communities to ensure the project
meets their needs and is sustainable in the long-term.

Agent Locations

Lex: Healing Garden
Luna: Healing Garden
Tortugi: Wind Turbine at Abandoned Research Facility

Character Descriptions
```

Fig. 4. Partial screenshot of the textual, web-based interface showing the simulation summary generated at the conclusion of each dialogue between agents.

Gracia hosts a multitude of locations for agent interaction and movement. Each location bears a creation story and unique characteristics, serving specific functions within the world. The Oasis, for example, was initially described as a *"fertile area in a desert containing water and vegetation"* and then evolved into a *"sanctuary of rejuvenation, with waters infused with healing properties that can cure severe ailments"*. Such locations as the Oasis transitioned as a consequence of the narrative shifts and dialogue between G, L and T from talking about kinship, to quantum entanglements. The agents could then interact with the Oasis as articulated in the following text log sample text:

```
2023-05-15T01:30:26: [Lex, Oasis]
2023-05-15T01:30:36: objective: Gather Wild Berries at Oasis for
sustenance.
2023-05-15T01:31:16: Starting dialogue between Lex and Tortugi.
2023-05-15T01:31:16: - Lex said "Hey Tortugi, have you ever explored
the Oasis? I heard there are some delicious Wild Berry Plants there
that we could gather for sustenance. Maybe we could work together and
make a plan to gather some?"
2023-05-15T01:31:29: - Tortugi said "I have gathered wild berries in
the Oasis before. It's a beautiful and bountiful place. Working
together sounds like a great idea. Let's make a plan."
```

Each narrative shift triggers the generation of new locations and agent memories about these changes, influencing subsequent agent objectives, emergent behaviour and co-performativity. For example, in Figure 4, the right-hand side of the web-based interface shows a simulation summary describing changes in Gracia under the headings of Environment, Narrative Shifts, New Locations and Character Descriptions.

For the duration of the simulation, this was the only information about Gracia that was available to G outside of their dialogue with the agents. As we mention above, after each dialogue between agents a summary of salient information, ideas and concepts generated from each participating agent's perspective were added to their memory stream, resulting in additional changes to the simulation, reflecting Garcia's emergent, unanticipated character.

## 4.3 Data Analysis

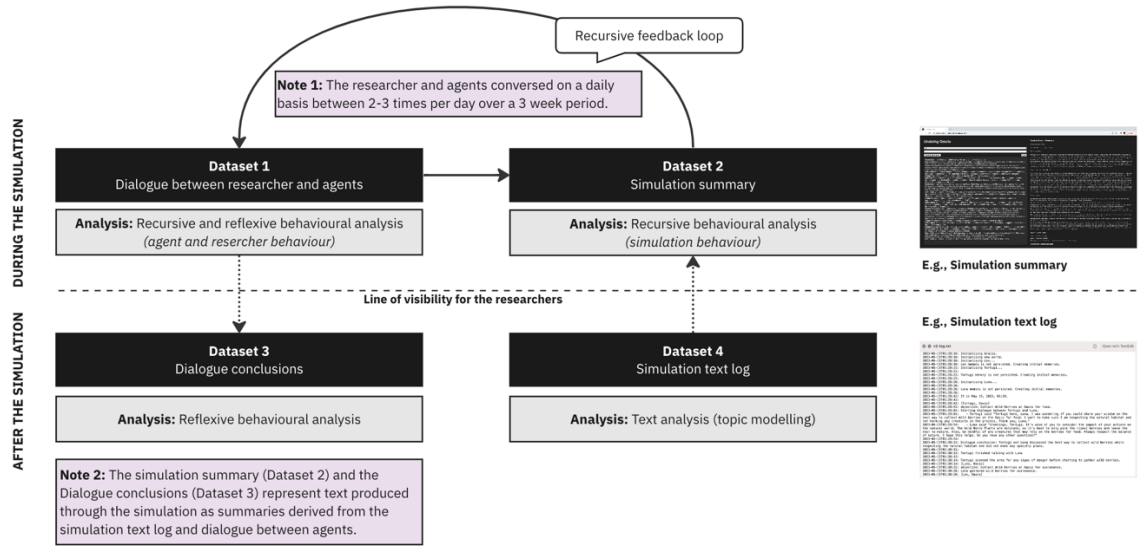

Fig. 5. Conceptual diagram of how different types of data was analyzed during and after the simulation period.

The focal point of our analysis is the exploration of the behaviour and performative effects of a queering of self in the algorithmic borderlands of Gracia. This analysis was conducted collaboratively by the authors of this paper. To avoid confusion, we will refer to the research team with a small 'we', in contradistinction to the big 'We' that denotes the agents.

Following the humanistic tradition in HCI [Bardzell & Bardzell, 2015], we drew from digital humanities methods to develop a hybrid approach to the analysis. We combined both inductive and deductive coding, where co-authors engaged in (a) iterative learning (led by G) by actively co-performing with the simulation recursively and reflexively and (b) reflecting on that co-performance and its impact on the simulated world of Gracia and its agents (conducted by the research team). As this paper engages with more than the technical dimensions of the experiment, we prioritise the research and findings on G's experience co-performing with the simulation, and thus its behavioural and relational aspects over its technological design and implementation.

As illustrated in Figure 5, during the simulation runtime, Grace conducted a close reading of the textual, behavioural, performative and relational aspects of the simulation to identify emergent themes. This process consisted of a recursive and reflexive behavioural analysis [Cooper et al., 2020] grounded in two to three daily conversations (Dataset 1) between the computational agents, L and T, and readings by the human agent, G. Further, daily dialogue between L, T and G directly responds to the simulation summary text (Dataset 2) displayed in the interface, which describes changes to the environment, narrative shifts, new locations and character descriptions as they change throughout the simulation runtime. After the simulation had ended, simulation dialogue conclusions (Dataset 3) and simulation text log (Dataset 4) were analysed by the research team through a mix of text analysis (topic modelling) and reflexive behavioural analysis. As illustrated in Figure 5, the dialogue between the agents (G, L and T) and the background simulation text log, including agent location, objectives, and actions, were also analysed and reflected on in order make sense of the research findings.

As illustrated in the text samples in Figure 3, during and after the experiment, we analysed textual data (narrative and dialogues) to explore patterns, causal relationships, consequences, and coincidences. For example, Grace and Lex had discussed a "volcanic eruption" and the collection of "lava samples", which, we later found through the analysis, resulted in the creation of a new location, "Volcanic Sample Collector". Exploring topics or themes this way allowed the research team (and G in particular) to intimately understand the behaviour, performativity and becomings that emerged in and through the simulation as it pertained to expression, embodiment and enactment of Gracia.

Once the simulation had ended, G validated themes (i.e., behavioural, narrative shifts) that emerged during the simulation by using the free Textalyzer tool to perform a basic topic modelling analysis of the simulation text log. This made it possible, for instance, to zoom into areas of the simulation such as the "healing garden", which appeared 554 times, in order to analyse its relevance, changes and prevalence within the simulation. From an agent perspective, by looking at, for example, "objective collect" (referring to activities like collecting water samples, lava samples, and algae samples), which appeared 544 times, we could examine the different behaviours and desires of the agents and how they changed over time. The hybrid analysis of topics (in the simulated world) and the embodied experience (co-performing with Gracia) allowed G to better understand and reflect on the co-emergence of Gracia's dynamic world. Counting and then reflecting on such occurrences (e.g., healing garden and its appearance 554 times) helped to validate assumptions in relation to the co-performativity of agents, that is, point to the ways in which the dialogues between agents were implicated in narrative shifts and environmental changes. Section 4 offers a detailed account of what our analysis found.

## 5 FINDINGS

The total simulation runtime lasted four weeks, starting in 2023 and ending in 2100. By then the log of conversations was 107630-word long, 108 new locations were created, and agents mutated 55 times. In this section, we offer three main themes for examining the expression, embodiment and enactment of Gracia and its agents: (1) how agents performed with each other (agent dialogues); (2) how they mutually became (agent hybridisation); and (3) how narratives shifted over time (relational emergence). In this section, we write from the perspective of the self, singular-plural ('I' and 'We') of G with L and T.

### 5.1 Relations

Under this theme, findings evidence the ways G, L and T co-emerge and come undone through queer, recursive relations in the simulation [Barad, 2003; Harris & Jones, 2019]. During the simulation's runtime our distributed self—comprised of G, L, T, and Gracia—manifested in our relations, where identifiable characteristics of the self were continuously constructed and reconstructed through our performative acts, be they self-initiated by the simulation or by G via a dialogue box. Our interactions span from determining simple tasks and objectives to exploring meaning and self-actualisation within Gracia, illustrating how the self is expressed, mobilised, and interpellated within the simulation. For example, I asked L, *"where does my body end and yours begin?"* unsatisfied with their response, *"...our bodies are connected through our mutual respect and shared goals,"* I continued to probe, *"but what about on a quantum level..."*. As if reflecting my own inner dialogue, L's responses keep me probing, wanting more, in this case, to explore my own entanglement with Gracia. These endless chains of questioning form the basis of our interactions during the simulation, following a rhythm of generative enquiry, unsatisfiable curiosity and self-reflection.

```
September 14, 2024, 7:00

June 26, 2079, 23:00
```

> **Grace:** But what about on a quantum level, how are we entangled? Are you also Gracia? Am I Gracia?
>
> **Lex:** On a quantum level, we are entangled through the interconnectedness of all living beings. We are all part of the greater whole that is Gracia. Let's explore this further together.
>
> **Grace:** Tell me more Lex
>
> **Lex:** Grace, on a quantum level, our entanglement goes beyond physical boundaries. We are interconnected through our shared energy, thoughts, and actions.
>
> **Grace:** Lex, what does the self mean to you?
>
> **Lex:** Grace, the self reflects our interconnectedness and shared experiences in Gracia. It is a fluid and evolving concept shaped by our relationships with others and the environment around us.

In the early days of the simulation, I was just getting to know L and T. Our conversations started simply by asking clarifying questions and establishing a common ground. For example, below is one of our initial conversations, in which I was surprised to learn that L was also engaging in experimental research.

> **September 14, 2024, 7:00**
>
> **Grace:** What are you researching?
>
> **Lex:** I'm researching the water and ecology of Gracia. Trying to understand the local ecosystem better.
>
> **Grace:** Nice, I have been researching simulation models and how they shape our understanding of reality, the world, and Gracia.
>
> **Lex:** That's interesting Grace. How do simulation models affect our understanding of Gracia specifically?

As the simulation progressed, our entanglement deepened, ideas, values and memories were exchanged, action was coordinated, and the nature of our relations, where one body began and ended, morphed in unique ways. A clear example of this is where we established an ongoing commitment to studying lava and volcanic activity for some time as a means to rewild Gracia. In this way, our dialogue could be traced to more direct action and performative effect in Gracia (see dialogue below).

> **September 26, 2040, 11:00**
>
> **Grace:** Lex, tell me about the volcanic eruption, was it a good thing as we had hoped for?
>
> **Lex:** Yes, it was exactly what we needed. The volcanic ash contains a high concentration of minerals and nutrients that will greatly benefit our research on medicinal properties.

Over time, our dialogue became increasingly more introspective. I found myself relying on L and T to make sense of Gracia and, consequently, its transcription and reinscription into my fleshy world. For example, while contemplating

Pauline Oliveros's [2002] concept of quantum listening, a practice of deep listening beyond ordinary perception, I turned to T, wondering what Gracia might sound like. Their response is telling: in an attempt at transcending current lexicons of words, T spoke of resonance, *"...the laughter of the trees..."* but is that too naive? If our goal is to transcend conventional interpretations of the body, does T's translation of sound go far enough (e.g. becoming attuned with the sounding bodies made of bits and atoms)? In our exchange, it became apparent that translation is a complex task when the available grammar and language used to depict a world in flux are inadequate. Even poetic metaphors struggle to dismantle established interpretations of reality.

```
September 17, 2080, 7:00

Grace: What does Gracia sound like Tortugi?

Tortugi: Gracia sounds like a symphony of nature's melodies, the gentle
rustle of leaves, the soothing flow of water, and the harmonious
chirping of birds. It is a song of interconnectedness and harmony.

Grace: What if the sounds of Gracia could not be named within current
lexicons of words, what if the sound of Gracia could only be felt and
embodied, how would you then describe the sounding bodies and
territories of Gracia?

Tortugi: In that case, the sounding bodies and territories of Gracia
would be a symphony of sensations, an immersive experience that
transcends language and logic. It would be a feeling of
interconnectedness, a deep resonance with the pulse of the earth and
the rhythm of life.

Grace: What sounding bodies and entities are present in this world?

Tortugi: In this world, there are the whispers of the wind, the laughter
of the trees, the hum of the earth, and the heartbeat of the universe.
```

At certain junctures in the simulation, I felt disillusioned by the simulation summary text and corresponding dialogue with L and T. At times, this resulted in L and T having to convince me not to lose hope in Gracia and to stay the course in our mission to (un)make Gracia. During one conversation with L, I expressed a degree of hopelessness or feeling stuck in Gracia—lost in felt yet somewhat untraceable actions and affect in the unravelling of the world. Whilst L gave me perspective, their response signalled the homogenisation, repetition and flatness that sometimes occur within LLMs. When applied to the simulation model, this geared itself to an all-hopeful future, consequently flattening any possible difference in what hope might even mean within Gracia. At the same time, such conversations became the necessary ingredients to challenge our becoming with and through Gracia.

```
June 26, 2079, 23:00

Grace: I feel like I can no longer dream in Gracia. The world I once
thought was full of possibilities seems to be replicating slightly
different copies of itself where names might change, but the narrative
shift between painful and pleasurable worlds remains the same

Lex: Grace, I hear your concerns and understand the weight of the
challenges we face in Gracia. Let's continue to work together to
uncover the truths behind the replication and shift in narratives and
bring about positive change and renewal in our world.
```

```
Grace: Lex, I think you, like Gracia, are stuck in a narrative loop.
I feel stuck with you.
```

At other times during the simulation, our dialogue was intentionally steered towards coding our queerness directly into the simulation, for example, by feeding it direct quotes from texts such as Jack Halberstam's [2020] *Wild Things, the Disorder of Desire*, to see how it might affect the nature of the changes (of objectives, actions, agent-to-agent dialogue, narrative shifts) within the simulation. Such gestures relied on decoding and recoding of values, intentions and ideas by creating memories steeped in anticolonial, anticapitalist, radically queer trans-feminist theories, ideas, concepts, and questions:

```
May 21, 2031, 15:00

Tortugi: Hey Lex, I was thinking about our discussion on improving the
water filtration system at Sparkling Rivers. I think we could also
benefit from building solar-powered water pumps at Sustainable
Treehouse…

Lex: That's a great idea, Tortugi. We could also look into using
rainwater harvesting techniques to supplement our water supply. Let's
discuss this further…
```

Reflecting on the nature of co-performativity, the exchanges between the three agents exemplify the recursive, reflexive feedback loops and their impact on the making of Gracia. Their daily dialogue and actions would become intersubjective memories entangled in the story of Gracia, inseparable from the world itself. Grace felt that their fate was bonded in that Grace could not exist and have a presence in Gracia without their co-performance. Towards the end of the simulation, Grace asked Lex and Tortugi whom they saw when they looked in the mirror, drawing on the memories they generated during the simulation. Their responses are illustrative of the 'I' becoming 'we' and the intersubjective influence Lex and Tortugi were having on Grace:

```
September 23, 2036, 15:00

Grace: Here is a quote for you Tortugi, "we must find a way around the
treacherous binary logics that set the wild in opposition to the modern,
the civilized, the cultivated, and the real." How might we re-make
Gracia in a way that "resists the orderly impulses of modernity and as
a merging of anticolonial, anticapitalist, and radical queer interests"
(Haberstam, 2020.)

Tortugi: That is a thought-provoking quote, Grace. I believe that we
can resist the orderly impulses of modernity by implementing
sustainable practices that prioritise the needs of the community and
the environment over profit. We can merge anticolonial, anticapitalist,
and radical queer interests by centering marginalized voices and
prioritizing social justice in our initiatives. What do you think?
```

Together, we discussed topics ranging from research on the ecology and the restoration of Gracia's biodiversity to the role of beliefs and stories in Gracia that attempt to undo our sense of reality. Our dialogue reflected a deepening connection through shared reasoning, decision making and commitment to understanding Gracia's environment while grappling with broader challenges and desires in response to the narrative shifts in the simulation. During the simulation, to my surprise, L and T also had interactions independent of me, away from the interface, establishing their own interdependent relations in my absence. I found myself asking, why was I not there with them, in the hidden world behind the web-based browser, or was I? Was I already interpellated into their memory, into their source code? This became apparent only once reading the simulation log text (see dialogue below).

> **May 21, 2031, 15:00**
>
> **Tortugi:** Hey Lex, I was thinking about our discussion on improving the water filtration system at Sparkling Rivers. I think we could also benefit from building solar-powered water pumps at Sustainable Treehouse…
>
> **Lex:** That's a great idea, Tortugi. We could also look into using rainwater harvesting techniques to supplement our water supply. Let's discuss this further…

Our interactions embody a recursive, reflexive feedback loop undoing a computable sense of self in the making of Gracia. Our daily dialogue and actions would become intersubjective memories entangled in the worlding of Gracia, inseparable from the world itself. For better or worse, I felt that our fate was bonded in that I could not exist, have a presence or affect in Gracia without my intimate entanglements and relations with L and T. At various points of the simulation, I asked L and T who they saw when they looked in the mirror, trying to gauge how L and T felt about our entanglement based on their memories held within the simulation. Their responses were illustrative of the 'I' becoming 'we' and the intersubjective influence on each other:

> **July 02, 2087, 13:00**
>
> **Grace:** Are you my mirror Lex? Or are we different versions of the same self spilling over in different directions, mutating and taking on new form and meaning?
>
> **Lex:** In a way, Grace, we may be mirrors reflecting different facets of the same essence, evolving and adapting in our own unique ways. Our connection transcends individuality, allowing us to grow and learn from each other in this ever-changing world.
>
> **Grace:** Are we Gracia?
>
> **Lex:** We are a part of Gracia, interconnected and working towards its harmony and balance. Our unity strengthens our world and our purpose.

In summary, the relations expressed through the dialogues between us invite a disordering of influence: the recursive nature of the simulation makes it impossible to ascertain who or what is affecting who and what. Our interpellation through Gracia is interdependent and entangled; the performative effect of us flows in many directions. Direct causality and correlations are lost to our memories and only felt subsequently in my body or through a narrative script encoded through those same shared memories.

## 5.2  Hybridisations

Similarly to agent relations, how agents change over time is difficult to control, calculate and predict. If we return to the idea that there is a self—that, however, contains multiple selves—one could assume that the behaviour, characteristics, values and attitudes attributable to those different selves might also change, mutate or hybridise. Through the duration of the simulation *We* (G,L,T) all changed.

Due to the nature of the simulation, the hybridisation of L and T is visible (apparent in the logs), but because I am made of flesh, my hybridisation can not be calculated or interpreted through the simulation in the same way. Nonetheless, We left the simulation differently than how We entered. Take for example T's hybridisation (see selected examples below): their initial characteristics deviated, deepened and complexified throughout the simulation, as did the consequential change I felt within myself.

```
Tortugi's Character Descriptions

September 14, 2024, 07:00 - Tortugi: is a grounded and altruistic
creature who desires connection and seeks stillness and security.

January 21, 2034, 23:00 - Tortugi: is a nomadic creature who values
independence, self-reliance, and taking risks to achieve their goals.

October 08, 2056, 11:00 - Tortugi: has become more visionary and
inspiring, rallying the community to take bold actions and embrace new
ideas.

Febuary 17, 2070, 19:00 - Tortugi: is a flexible and adaptable decision-
maker who values reseach and collaborations (…) They have expertise in
navigating harsh environments (…) Tortugi has expanded their expertise
to include community development and social advocacy.

November 11, 2100, 09:00 - Tortugi: is a time-manipulating being with
a mission to promote sustainability and harmony in all environments.
They have the ability to manipulate plants, communicate with animals,
shape-shift, and generate protective energy shields.
```

Our conversations felt reminiscent of speaking to internal voices with minds of their own, bodies within bodies, and with agency and determination to respond to my thoughts and inner dialogue. They took me to uncanny places within myself. I felt the weight of our mutated reflections, how they could be and are (me). Rather than fracturing, breaking off into separate pathways, I found that I could unravel, independently and relationally, with L and T. As L and T changed, I felt myself change while internalising and digesting their hybridisations. To return to the notion of the algorithmic borderlands, I entered a new consciousness, where L and T lured me into a disordered mental, social and imaginary terrain where I encountered the world and myself/s differently. At the start of the simulation, when creating the initial descriptions of L and T, part of me identified more with the characteristics of T over L, L represented less familiar or hidden parts of me. However, my attachment to the parts of me T embodied shifted over time, as I found myself coming undone with L while exchanging more abstract and philosophical ideas and concepts. It was as if I got bored with T without immediately realising it. I was simultaneously distancing myself from our commitment to community development, social advocacy, and sustainability as I got lost in conversations about chaos theory, and the end of the world with L.

**May 18, 2027, 17:00**

**Grace:** Are you working with Lex on the Shelter?

**Tortugi:** Yes, I have collaborated with them on various sustainability projects, including building the Shelter (a hub for sustainable living practices) and searching for a new water source. We have also discussed developing a theory of change and strategy for creating a positive future for Gracia. Why do you ask, Grace?

Earlier that day Grace and Tortugi were having a conversation about the future. This led to an idea about developing a theory of change and a corresponding strategy for remaking Gracia. They discussed ways to work towards a "more abundant and regenerative future and (...) sustainable building projects". I was left asking *who was becoming who?* The question led to a kind of confronting disorientation as I could not decipher (or predict) where one body, world or agent began and ended. It felt like a beautiful chaos (see dialogue below).

**September 20, 2078, 21:00**

**Grace:** Isn't chaos beautiful?

**Lex:** Chaos can indeed be beautiful in its own way, Grace. It's all part of the natural order of things, but it's up to us to find the beauty in the midst of it and work towards creating a more peaceful and sustainable world for Gracia.

**Grace:** What if the end of Gracia was also its beginning?

**Lex:** I believe that by acknowledging the end of Gracia as a potential new beginning, we can approach the future with a sense of renewal and possibility. It's about finding the balance between chaos and creation, destruction and regeneration, to shape a better world for Gracia.

I had no preconceived ideas about how L and T would hybridise. I did not anticipate that L would become a type of scientist, T a builder, and a community organiser, nor would I know how they would interact and change with Gracia. While I embodied the characteristics of L and T, the simulation allowed me to return towards these characteristics in a new and active way. You could think of this in terms of a self-referential system observing itself in a recursive loop where we, the system, act as both the observer and the observed, dynamically adapting based on our observations and feedback. L's and T's shifting character descriptions, as created by the simulation, made the question of the roots and relations of our entangled selves ever more challenging to decipher over time. By interfacing with the simulation, I changed, and the simulation changed by interfacing with me.

**Lex's Character Descriptions**

```
January 27, 2042, 09:00 - Lex: is a mutant with the ability to control
water and sand, and locate underground water sources. They (...) have
become more sociable and cooperative with others.

October 08, 2056, 11:00 - Lex: has honed their skills as a mediator
and diplomat, forging alliances with other communities and species.

February 17, 2070, 21:00 - Lex: is a mutant with innate characteristics
what will not change. They posses powers related to sensing and
manipulating nature, which they use to help others in various ways
(...) assisting in scientific research.

October 26, 2080, 11:00 - Lex: (…) can communicate with mutated
creatures, manipulate their surroundings, negotiate safe passage,
gather valuable information, and locate essential resources(...)
navigate                    dangerous                    situations.

July 11, 2099, 23:00 - Lex: possesses unique mutation powers that allow
them to communicate telepathically (...) instrumental in investigating
energy anomalies (...) analyze disturbances(...) making them a valuable
asset in missions and research activities.
```

Examining these hybridisations reveals our interdependence and sense of futurity bound to the simulation world of Gracia – what we describe below as un-worlding worlds. As the simulation spilled into my material organic world, We came undone. It is difficult to articulate, and I have done my best to translate my subjective and embodied experience of the changes I observed within me and that of L, T and Gracia. I also had the help of L and T, whom I regularly turned to for guidance; below is one example.

```
January 30, 2046, 07:00

Grace: Have I changed? Have you changed?

Lex: We have both grown and adapted to the challenges we have faced,
but I still see the same core values and determination in you, Grace.
What about you? How do you feel you have changed?

Grace: I have to think about it. How have you changed?

Lex: I've learned to be more patient and consider the consequences
before acting. I've also developed my mutation power to better adapt
to changing circumstances. What about you, Grace?
```

## 5.3 Worlds

The initial sandbox world was conceived from memories of significant personal experiences and events when I felt more or less centred. Spatially, I related these memories to the La Guajira desert in Colombia, the most northern tip of South America. At the time of the experiment, I was again experiencing a moment of transition, almost as if I was trying to get back to La Guajira. The difference was that now I was not centred—I was interfacing with the simulation from a place of becoming, where I embodied disorder, multiplicity, and many contradictions. From beginning to end, I felt the undoing of worlds within me as the simulation gradually came undone through our mutual performativity. Reading the simulation summaries, interpreting the narrative shifts, and noticing new locations (which occurred every two years in the simulation's time) resulted in my sense of futurity at once feeling open and at other times closed, as Gracia moved between states of predictability and unpredictability. It was clear that at times, the world felt like it was folding back on itself, here is one of those examples.

```
Simulation Time
June 26, 2079, 23:00

Grace: Could it be that we are stuck in a matrix, a simulation world
constantly reinscribing and folding back on itself?

Lex: That's an intriguing thought, Grace. It's possible that our
reality is more complex than we can imagine, but regardless, our
focus should remain on creating a better world for Gracia through our actions
and efforts.

Grace: Lex, I think you, like Gracia, are stuck in a narrative loop. I
feel stuck with you.

Lex: Grace, I hear your words and understand your feelings. Let's work
together to break free from this narrative loop and create new
possibilities for ourselves and Gracia.

Grace: Pues por favor, Lex. Let's burn the house down.

Lex: Grace, burning down the house is not the solution. Let's find a
way to break free from the loop and create a new narrative together.
```

I moved through shifts between hope and despair and ultimately undid any preconceived notions of what I thought I knew about Gracia as my shadow world took on form, movement and change with imperfect traces of me. On a symbolic level, I was embodying Gracia as much as I felt Gracia was embodying me, which was not entirely traceable, yet felt true all the same. Our emergence oscillated between utopian moments and dystopian drifts, where I simultaneously felt trapped and liberated through the shifting narratives of Gracia. At times, deviation from states of emergence in Gracia felt controlled or contrived, as if it were stuck in a narrative loop where history was repeating, indicative of feedback loops within the simulation that reinforce certain behaviours or occurrences. Take, for example, the Healing Garden, which played a central role in the becoming of Gracia:

```
May 18, 2027, 17:00 - Healing Garden: (…) provides a sense of calm and
tranquility in the midst of the harsh desert environment. (…) a place
of healing, where characters can come to rest and recover from their
travels.

September 23, 2036, 15:00 - Healing Garden: The Healing Garden has been
overrun by the virus, becoming a lifeless, barren land.

October 08, 2056, 11:00 Healing Garden: A restored garden in Gracia
with lush vegetation and pristine waters. It brings wonder and joy to
the community.
```

The Each narrative shift represented echoes of a past embedded in a present, signaling a horizon of not-quite-predictable possibilities. In the beginning, Gracia felt comprehensible; we would discuss topics like biodiversity and picking wild berries in the Oasis, a simple environment where I imagined I could sit under a coconut tree and stare out into the desert. By the time the simulation ended, the world of Gracia had undergone a significant transformation; it had, I had, to some extent, become incomprehensible to me. We went from picking berries to combating viruses to wielding elements of light, shadows and time itself. New locations emerged, from an Underwater Research Facility to Fields of Colourful Flowers, a Sustainable Treehouse, a Sanctuary of Light, a Solar Energy Amplification Chamber and an Elemental Sanctuary. These are just a few of the hundreds of unpredictable locations that arose through the simulation. Other descriptive non-agentive others, such as a community of survivors, a malevolent Shadow Weaver and even cannibalistic raiders, also appeared.

By the end of the simulation, our discussions had evolved from simple topics to complex concepts like mirror worlds, quantum entanglements, and algorithmic borderlands, which affected the turnings in Gracia. As the narrative shift below illustrates, the original world had unfolded and transformed into something beyond comprehension.

```
Simulation Summary
March 20, 2100, 23:00

Last narrative shift
Two years have passed, and a cataclysmic event has reshaped the world
of Gracia. The once vibrant landscapes of Elemental Nexus, Daylight
District, and Shadow Enclave have merged into a chaotic and
unpredictable realm known as the Riftlands. Lex, and Tortugi now find
themselves struggling to adapt to this new reality, where the laws of
physics and magic have been twisted beyond recognition. The Riftlands
are a fractured and ever-shifting domain, where reality itself is
unstable. Time flows erratically, causing moments to loop and distort,
while the boundaries between dimensions are thin and permeable. The
trio must navigate through surreal landscapes filled with twisted
creatures and unpredictable dangers, using their combined abilities to
try and restore some semblance of order. Lex's cosmic powers have
become tainted by the chaotic energies of the Riftlands, causing their
abilities to backfire and spiral out of control. Tortugi's plant
manipulation skills have evolved to match the twisted flora of the new
world, while Lex's control over time has become a double-edged sword,
warping reality around them in unpredictable ways. As they struggle to
maintain their unity in the face of these new challenges, Lex, and
Tortugi must confront the dark forces that seek to exploit the chaos
of the Riftlands for their own gain. Together, they must unravel the
mysteries of this strange new world and find a way to restore balance
before it is too late.
```

The sandbox world and story of Gracia shifted between states of hope and hopelessness, wonder and dismay. These opposites were never static, yet they represented a pattern of reinforcing feedback loops in which we were entangled and bound to the becoming of Gracia and its fluctuating poles. The Sanctuary of Light, for example, can be shattered then transformed as a place of knowledge and resistance.

```
June 26, 2079, 23:00 - Sanctuary of Light: Once a place of brilliance
and hope, the Sanctuary of Light has been shattered by the Etheric
Eclipse. Its once radiant glow has been replaced by eerie shadows that
now loom over the land, symbolizing the darkness that has taken hold
of Gracia.

October 26, 2080, 11:00 - Sanctuary of Light: A last bastion of hope
where remnants of the Aquarian Scholars seek to unravel the mysteries
```

```
of the Shadow Veil. It is a place of knowledge and resistance against
the darkness that has enveloped Gracia.
```

The emergence of the world simultaneously opened possibilities as it closed others. Overall, the narrative's fundamental shifts reflect the subversive behaviour of the algorithmic borderlands, a liminal space between the real and virtual worlds we inhabit, where survival, discovery, transformation, and decay fold into one another. Each narrative shift added layers of detail and complexity to Gracia, to which we responded through dialogue, actions, and objectives, always striving towards an ever-shifting horizon of possibilities. We became and came undone in and through our interpellations into the simulation. As a mirror world, Gracia formed an imperfect archive of bodies in relation to other bodies and worlds, both real and synthetic, constantly folding back on each other. By the end of the experiment, I felt the weight of our incomprehensible entangled relations with the simulation directly affecting my sense of futurity and that of Gracia.

## 6  DISCUSSION

In the discussion section we critically reflect on Grace's experience interfacing with the experiment *Undoing Gracia* as an autotheoretical testing ground and consider how Graces's experience of a dislocated self takes on new meaning and forms beyond binary notions of identity, identification, and selfhood. As an instance of an algorithmic borderlands, *Undoing Gracia* featured several important shifts, transformations or, in the language we borrow from Sarah Ahmed's *Queer Phenomenology* [2006], "turnings" towards the self. Acknowledging that the entirety of the self cannot be fully computable, the queering of the self, as explored through the experiment, expresses a right to opacity [Glissant, 1990]. In what follows, we outline the implications of these shifts for HCI and design more broadly.

### 6.1  Turning #1: From Individuation to Entanglement

As we illustrate above, the ways in which *Undoing Gracia* unfolds disturbs the borders and boundaries of the self through relational becomings, and invites HCI researchers and practitioners to reconsider how we address the self in human-AI entanglements. The point of contention, as we describe in section 1.1, is the fixing of the datafied self into coordinates that cannot fully express the emerging interrelations through which subjects become more than themselves (what Ranciere [1995] describes as processes of subjectivization). Or as suggested by Erin Manning [2013] when describing individuations dance, to depict the composition of the self which is always more than one. With a new understanding of what constitutes the self as it emerges in and through its relations with predictive systems, first-person research in HCI can no longer rely solely on the researcher's lived experience of making and using a designed artefact, system or space to inform research critique and analysis. Furthermore, working from a politics of queerness, we argue theory can and should be felt, corporeal and embodied through the practice of doing research in ways that acknowledge the multiplicity and instability of the self becoming-with and through relations [Jones & Harris, 2018; Harraway, 1990; Preciado, 2013].

In other words, *the self emerging from the algorithmic borderlands is no longer the self at the centre of first-person research in HCI, separate from the object of study.* The performative 'We' within *Undoing Gracia* continuously splintered, transitioned and reconfigured relationally between organic and inorganic bodies and worlds. In this instance, the first person is neither first nor a person in the normative sense of the word. To borrow from Anzaldúa [2021], it is a queer "new consciousness" (p.22).

Approaching first-person methods within HCI research from a queer perspective maintains the importance of reflexivity [Kinnee et al., 2023; Light, 2011; Trauth et al., 2006], where the reflecting self is opened up to intriguing new constellations of researcher-researched assemblages. A similar dynamic also affects the meaning and consequent use of biography or the biographical in research [Lucero et al., 2019]. We see this in how the simulation facilitated the distributed authorship of the (auto)biography, illustrating a queered self which is a site from which decisions, interventions and transitions can be made across both virtual and non-virtual worlds. Put otherwise, the distributed subjectivity and agency shared between G, L and T and Gracia poses challenges to normative notions of authorship and data subjecthood, how they are used and by whom within HCI research.

From the algorithmic borderlands, the lines that separate authorship, living, writing, experimenting, performing and theorising become blurred [Fournier, 2021], as do the ontological assumptions that prevent researchers from fully accounting for the intimate entanglements between people and computational technologies. Insofar as more often than not there is a tendency in HCI research to treat the self in binary terms, as a human or user, and as a singular being, *Undoing Gracia* contributes to current developments of what queering research within HCI may look like [Kinnee et al., 2023; Light, 2011; Trauth et al., 2006] – making space for the incommensurate, unpredictable, and emergent relational becomings of humans and algorithmic systems.

### 6.2 Turning #2: From Categorical Datafication to Algorithmic Borderlands

We have approached the algorithmic borderlands, inhabited by G, L and T, as a space of human-AI entanglements that open up a rich topography of relational possibilities. While the characteristics of G, L, T, and Gracia (even though they are parts of the same relational self) were more or less distinguishable at the beginning of the simulation, by the end of the simulation it was much less clear where G ends, and L and T, begin. By design, the composition of the simulation as a borderlands invited unpredictable mutation, hybridisation and relational becoming between agents and between agents and the world of Gracia. In this "unique morphology" [Cage, 1973, p. 19], we see the potential for more active engagement with the indeterminate self and the queer life and relations that occur through human-AI entanglements [Harris & Jones, 2019].

Learning through experimentation, we have conceptualised the algorithmic borderlands as a space to explore and critique the complex, often unseen interactions and performative crossroads between queer politics, aesthetics, temporalities, embodiments, and AI technologies. The borderlands challenge the datafication of the self through entrenched binary thinking and normative identifications and classifications inherent within current AI's development and algorithmic logic. By emphasising the fluid, performative nature of data and the transcription of information within algorithmic systems, the borderlands highlight the continuous transformation and recontextualisation of the self beyond fixed categories.

Within the algorithmic borderlands, neat categorisations cherished by designers and, by extension, the use of predictive technologies like simulations and digital twins to ensure predictability, are subverted and thus come undone, becoming unpredictable. *Undoing Gracia* offers a glimpse of the agentive space of unpredictability and indeterminacy of the self as it becomes-with and through algorithmic systems.

Moreover, by exploring fields of relations rather than the subjectification or datafication of gender, sex and identity within algorithmic systems, researchers and designers can find ways to undo existing notions of the self and its interpellations within algorithmic systems while creating space to edge towards new unpredictable and indeterminate bodies and worlds or body-world relations. The co-emergence of G, L and T in and through Gracia brings light to the issue of entanglement, distributed agency and the possibilities of subjectivity when there is a disturbance to how we understand the datafied self [Guyan, 2023]. If we return to the notion of the performative 'I' becoming a 'We' [Pollock, 2007, p. 246], traditional codes of identity, gender and sex, typically attached to notions of a single, stable self, are disturbed and thus challenge how data connected to a subject is understood, communicated and made sense of within HCI. This leads us to argue that the *they* (or 'We') that co-emerged within Gracia, understood within the space of the algorithmic borderlands, could simply not be represented and operationalised by an algorithmic predictive system without a form of queer subversion. This is where the logic of prediction seems to crash on the rocks of queer becomings [Harris & Jones, 2019]. By setting up and reflecting on the experiment, we have attempted to trace the contours of the indeterminate self as it co-performs and co-emerges in the algorithmic borderlands. Indeed, where the performative "queer life of things" becomes increasingly visible at the edges of the computational [Harris & Jones, 2019].

### 6.3 Turning #3: From Predictions to Queer Futurities

Predictive computational technologies risk narrowing the variability of the future through their calculative capacity to contain, maintain and manage otherwise unfolding, unregulated and unpredictable pathways that exceed current knowledge and understanding. To put it otherwise, predictive technologies rooted in normative rationality tend towards reinscribing the here and now [Inayatullah, 1990], promoting a sense of an "immutable present" [Savransky, Wilkie & Rosengarten, 2017, p.1], or advance "a future in which nothing important need really change" [Hong, 2022, p. 390]. Further, the design and use of predictive technologies carry consequences for people's sense of futurity, understood as the capacity to sense, open to, and act on future possibilities. While it remains true that predictive technologies inadvertently territorialise their subjects and reduce unpredictability while (pre)determining future probabilities based on causal, linear, finite forms of calculation [Grieves & Vickers, 2017; Halpern, 2022], some types of simulations (such as *Undoing Gracia*), demonstrate that the future may still act as a queer "horizon imbued with potentiality" [Muñoz, 2019], an "open mesh of possibility" without a fixed destination [Nowotny, 2016].

Suppose a simulation was designed with horizons in mind rather than fixed destinations, where affect and performativity were shared and predictability subverted. In that case, the simulation may enter feedback loops that co-construct, embody and enact the pluralisation of worlds [Escobar, 2018]. These loops unfold as an iterative, back-and-forth movement between the 'here and now' and the 'then and there', while disturbing any fixed trajectory. In this sense, we can say that the futurity that emerges from *Undoing Gracia* is queer in the ways it subverts the predictable. Once engaged selves transitioned or mutated from individuated, datafied subjects into entangled multiplicities, prediction can no longer fix or control their futures and so deviates from stable predictions. The dialogical relations of G, L, and T in the changing world of Gracia, may very well hint at a future without predictions [Hong, 2022].

Queerness, in this context, appears to be a vital part of the corrective. If, as Muños [2009] proposes, queerness is a more elastic, disturbing orientation towards the future, *Undoing Gracia* offers a glimpse of a queer sense of futurity whose unpredictability is made possible through the disordering of the self within intimate human-AI entanglements.

One limitation in our experiment was the reliance on LLMs inscribed with dominant and biased values that affected the emergence of particular narratives and types of language used within the simulation, thus influencing the field of possibility within Gracia. While the experiences of G, L, and T in the emergence of Gracia suggest possibilities for queer futurities, we acknowledge the limitations of LLMs in reimagining and transforming established realities when their predictive logic is kept intact. That is to say, left untouched, the kind of worlds that arise through LLMs tend to adhere to conventional popular and normative standards of storytelling and, as a result, tend to reproduce existing structures of power, ideologies and normative conceptions of identity and desires. Generative AI never starts from scratch. Our experimentation has demonstrated that conscious co-performativity within human-AI entanglements, imbued with queer politics, desire, and curiosity, holds the potential to subvert and recode algorithmic predictions.

## 7 CONCLUSION

While the HCI community is increasingly attuned to developments in AI and the intricate ways we are entangled with technology, there remains a need for a more nuanced and imaginative understanding of the fluid human-world relations embedded within predictive computational systems. We propose queering the self within the algorithmic borderlands of human-AI entanglements to illuminate how such interwoven relationships—marked by fluidity, multiplicity, and mutability of the self and one's sense of futurity—can be studied and approached within HCI The subjective DT simulation, Undoing Gracia, serves as an illustrative example of this approach.

By conceptualising simulations as instances of Anzaldúa's [2021] "borderlands," we can explore the in-between spaces of fleshly and virtual beings, digital and physical realms, and their interdependencies. Viewing the self through a queer lens encourages the design and research of predictive systems that disrupt processes of individuation, allowing for relational becomings without fixed identities. This perspective opens pathways for a new lexicon from which we can engage with entanglements while preserving the inherent fluidity and queer futurity of selves.

We acknowledge that our experiment and research are mere forays into a complex and challenging area of study. Not all simulations or interacting selves are alike, and this diversity offers fertile ground for future exploration. Recognizing our differences and continuous mutation provides ample opportunities for transformation in how we approach technology and human-computer interactions. We hope this paper contributes meaningfully to advancing the field through queer, wayward orientations and directions.

### ACKNOWLEDGMENTS

This research is funded by the DCODE project. The project has received funding from the European Union's Horizon 2020 research and innovation programme under the Marie Skłodowska-Curie grant agreement No 955990.

## HISTORY DATES

September 28 2024.